\documentclass[a4paper]{jpconf}
\usepackage{graphicx}

\usepackage{color}
\usepackage{amsmath}
\usepackage{bm}
\usepackage{amsfonts}
\usepackage{epsfig}
 \usepackage{graphicx}
 \usepackage[bbgreekl]{mathbbol}
 \usepackage{dsfont}
 \usepackage[utf8]{inputenc}
  \usepackage[T1]{fontenc}
  \usepackage[normalem]{ulem}

\begin{document}
\title{Screening of a single impurity and Friedel oscillations in Fermi liquids}

\author{Banhi Chatterjee and Krzysztof Byczuk}

\address{Institute of Theoretical Physics, Faculty of Physics, University of Warsaw, ul.Pasteura 5, PL-02-093, Warsaw, Poland}

\ead{Banhi.Chatterjee@fuw.edu.pl}

\begin{abstract}
	 We numerically study Friedel Oscillations and screening effect around a single impurity in one- and two-dimensional 
	 interacting lattice electrons. The interaction between electrons is accounted for by using a momentum independent self-energy obeying the Luttinger theorem.
	 It is observed in one-dimensional systems that the amplitude of oscillations is systematically damped with increasing the interaction while the period remains unchanged.
	 The variation of screening charge with the impurity potential is discussed. We see that the screening charge is suppressed by the interactions. 
	 In case of two-dimensional systems the surface oscillations around the impurity  are more localized with increasing the interactions.
\end{abstract}

  \section{Introduction}
		The screening of impurities and density oscillations are characteristic phenomena for quantum many-body fermionic systems. These ripples are known as 
		Friedel Oscillations (FO) \cite {friedel1952xiv}. Classically no ripples are observed around the impurity. 
		The finite wavelength of the electron is attributed as the cause of the FO. Several experimental studies have been conducted on FO. 
		E.g., in 1981 Binnig \textit{et al.} \cite{PhysRevLett.49.57} observed FO  using Scanning Tunnel Microscopy (STM) \cite{guntherodt_scanning_1994}. In 1990 Eigler 
		and Schweizer observed FO in quantum corrals on Cu(111) surface at 4K \cite {eigler_positioning_1990}. Hasegawa \textit{et al}. observed FO at 5K on 
		Si(111)Ag surface \cite{hasegawa_real-space_2007}. Most of the theoretical and experimental studies have been conducted for non-interacting or 
		weakly interacting systems till now \cite{simion2005friedel,PhysRevLett.75.3505,PhysRevB.78.195124,grosu_friedel_2008,kroha_correlation-enhanced_????}.
		An exception is the study of one-dimensional Luttinger liquid where a strong renormalisation of FO has been found theoretically by Matveev \textit{et al.} \cite{matveev_scattering_1995}.
		Thus it is of potential interest to study further the phenomenon of FO in correlated electronic systems in higher dimensions where Landau quasiparticles exist \cite{nozieres_theory_1997}.
		In this paper we theoretically address the question if interaction between electrons affects the amplitude, the period, and the phase of FO, and how.
		We conduct our study for 1D and 2D systems assuming that they are in a Fermi liquid state which allows us to investigate the influence of dimensionality on FO.

  \section{Model and Formalism}

		We consider a lattice system with the one-particle Hamiltonian
	
		    \begin{equation}
			H_{0}= \sum_{ij} t_{ij}\  a_{i}^\dag\ a_{j} +\sum_{i}V_{i}\ a_{i}^\dag\ a_{i}  , 
		     \end{equation}
		where $a_{i}$ ($a_{i}^{\dag}$) is the annihilation (creation) fermionic operator on the $i^{th}$ site and $t_{ij}$ is the hopping matrix element between the $i^{th}$
		and $j^{th}$ site. The inhomogeneous potential is $V_{i}$. We will focus on one-dimensional (1D) and two-dimensional (2D) lattices in the presence of a single site impurity, which is modelled
		by assuming that $V_{i}=V_{0}\delta_{i i_{imp}}$, where $i_{imp}$ is the impurity site. The nearest neighbour hopping $t=1$ sets the energy units.\\
		When the interaction between the electrons is present this Hamiltonian must be extended by adding the corresponding interaction term $H_{int}$. The one-particle properties 
		are found from the one-particle Green's Function $G(\omega)$ in the frequency $\omega$ space, which in the real-space (lattice) representation
		obeys the matrix Dyson equation \cite {rickayzen1980green,doniach_greens_1998}.
      
		    \begin{equation}
			\mathbb{G}(\omega)= [\mathbb{G}_{0}^{-1}-\mathbb{\Sigma}(\omega)]^{-1} ,
		    \end{equation}
		    where 
	         \begin{equation}
		  \mathbb{G}_{0}(\omega)= [(\omega+\mu)\mathbb{I}-\mathbb{H}_{0}]^{-1}  ,
		 \end{equation}
		 and $\mathbb{\Sigma}(\omega)$ is a self-energy matrix which takes into account all interaction effects. The non-interacting Green's Function 
		 $\mathbb{G}_{0}(\omega)$ is determined from the Hamiltonian (1) in a matrix representation $\mathbb{H}_{0}$. The chemical potential is denoted by $\mu$.\\
		 
		 In accordance with the Real Space Dynamical Mean-Field Theory (R-DMFT) approximation we assume that the self-energy matrix is diagonal, i.e.
		 $\Sigma_{ij}(\omega)=\Sigma_{i}(\omega)\delta_{ij}$ \cite{snoek2008antiferromagnetic,vollhardt2012dynamical,PhysRevB.59.2549,PhysRevLett.100.056403,freericks2006transport}. 
		 Solving the full R-DMFT self-consistent equations is computationally exhaustive. Therefore in order to have an initial insight about FO in interacting systems we consider
		 only the model self-energy obeying the Luttinger Theorem in the following form, 
		 
		 \begin{equation}
		    \Sigma(\omega)=g \frac{a\ \omega}{(\omega+i\ a)}\frac{b}{(\omega+i\ b)}  ,
		  \end{equation}
		 where a, b, g are real, positive parameters \cite {inoue1995systematic,byczuk_phenomenological_2002}.
		 The electronic correlations in our system is varied by tuning the parameter g. The model self-energy has proper Fermi liquid properties at small $\omega$ and correct assymptotics
		 at large $\omega$.\\

		The FO in the interacting systems are investigated by computing the local spectral function 
		\begin{equation}
		   A_{i}(\omega)=-\frac{1}{\pi}\text{Im}\ G_{ii}(\omega)    ,
		\end{equation}
             where we used the analytical continuation $\omega \rightarrow \omega+i0^{+}$.
	     Finally, we compute the local site occupation from
		\begin{equation}
		{n}_{i}=\int_{-\infty}^{+\infty} A_{i}(\omega)f(\omega) \,d\omega   ,
		\end{equation} where $f(\omega)=1/(1+\exp(\beta\omega))$ is the Fermi-Dirac distribution function and $\beta=1/kT$ is the inverse of the temperature. The chemical potential
		$\mu$ is fixed by preserving the average density of particles  
		
		\begin{equation}
		 \bar{n}=\frac{1}{N_{L}}\sum_{i=1}^{N_L} {n}_{i}  ,
		\end{equation}
		where $N_{L}$ is the number of lattice sites.

	    \section{Results and Discussions}
		   
		   The local occupation (6) is determined by inverting numerically the equation (2) for the Green's Function. We used the Periodic Boundary Condition (PBC) for a finite 1D and 2D
		   lattices. The inverse of the temperature $\beta=10000$ in all cases and $\mu$ corresponds to the half-filled band.

			\subsection{1D interacting system}
			
\begin{figure}[h]
                 \includegraphics[width=25pc]{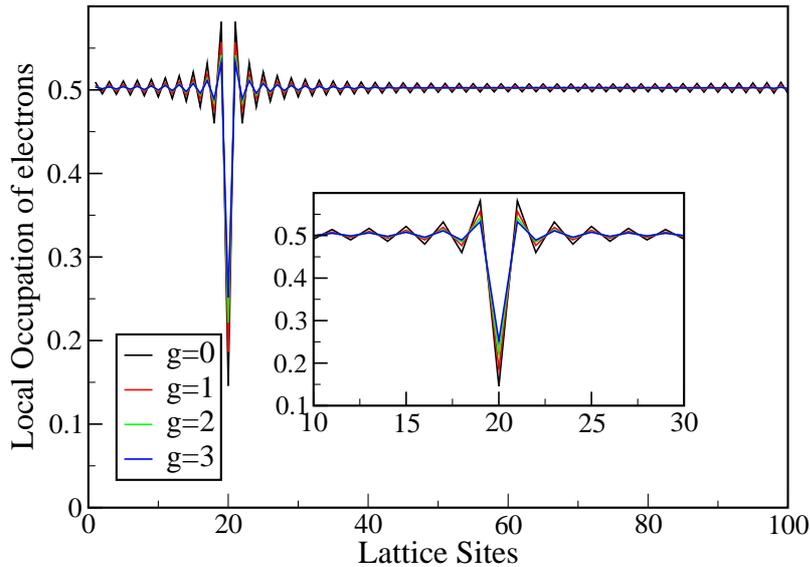}\hspace{2pc}%
                 \begin{minipage}[b]{14pc}\caption{\label{fig:fomag}Variation of the local occupation and FO in presence of the single impurity potential
                 $V_{0}=2.0$ placed at $20^{th}$ site in a 1D lattice chain. Different colours correspond to different electronic interactions. The inset shows behaviour of FO 
                 in the neighbourhood of the impurity site.}
                 \end{minipage}
                 \end{figure}

			In Fig. \ref{fig:fomag} we present the local occupation $n_{i}$ in a one-dimensional lattice with $N_{L}=50$ sites where the impurity is located at the $20^{th}$ site. 
			The repulsive potential is $V_{0}=2$. The particles are repelled from the impurity site and the screening cloud is built around it. We can clearly see the characteristic FO.
			The screening cloud is seen for all g parameters, however, the amplitudes of the density oscillations is suppressed. The period of oscillations remains unchanged. We obtained
			similar pictures for other values of $V_{0}$ and positions of the impurity.\\
			
                 \begin{figure}[h]
                 \includegraphics[width=20pc]{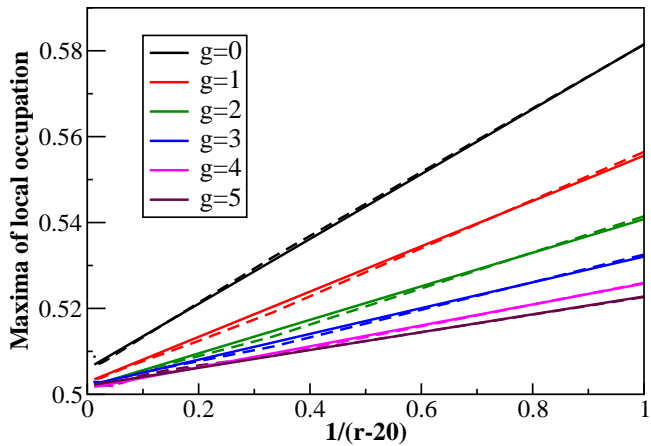}\hspace{2pc}%
                 \begin{minipage}[b]{14pc}\caption{\label{fig:fodecay}The maxima of local occupations decay inversely with the distance from the impurity site, i.e. $A/|r-20|$. 
                 Solid lines show the data while dotted lines show  linear fits.}
                 \end{minipage}
                 \end{figure}

\begin{figure}[h]
                 \includegraphics[width=20pc]{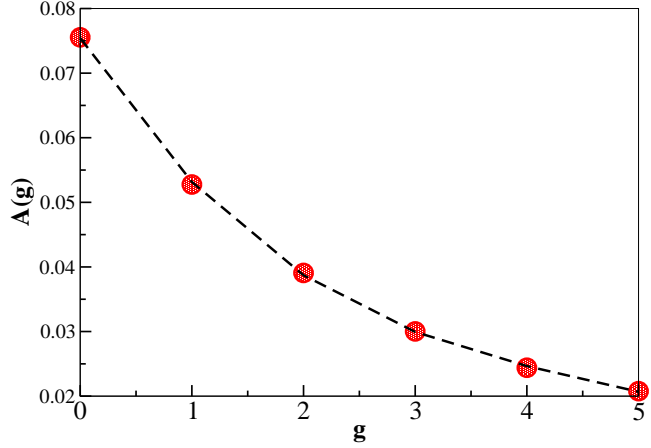}\hspace{2pc}%
                 \begin{minipage}[b]{14pc}\caption{\label{fig:slopeg}The amplitude of oscillations varies as a decreasing function of g.}
                 \end{minipage}
                 \end{figure}

		In the non-interacting case in a continuum the density of particles is assymptotically described by
			
			\begin{equation}
			 n(r)=n(0)+A \frac {cos(2k_{F}r+\delta)}{r} ,
			\end{equation}
			 where 
		$n(0)$ is the uniform density, A is the amplitude of FO, $\delta$ is the phase shift of the scattered states, $k_{F}$ is the Fermi momentum, and r 
		measures the relative distance from the impurity. Since for $D=1$ the Fermi vector $k_{F}$ is unchanged with using the k-independent self-energy (4) the period of oscillations must be the same 
		for different g. In addition the phase $\delta$ is unchanged as well since we impose the PBC. As seen in Fig. \ref{fig:fodecay} the way how the oscillations
		decay is described by the $1/r$ law. Due to the interaction only the prefactor A behaves as a decreasing function of g as shown in Fig. \ref{fig:slopeg}. \\

                 \begin{figure}[h]
                 \includegraphics[width=20pc]{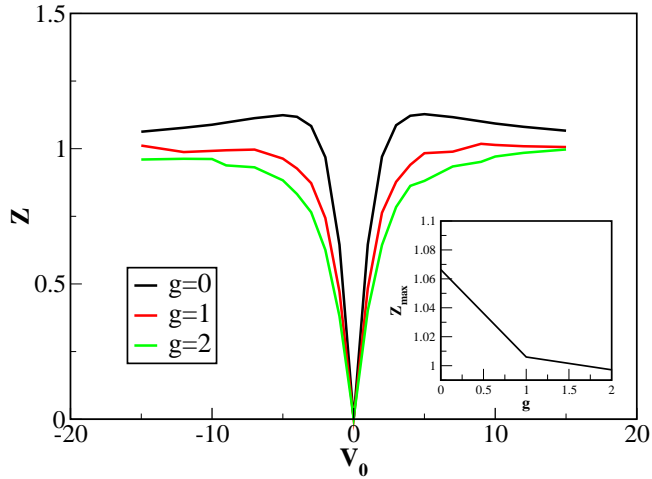}\hspace{2pc}%
                 \begin{minipage}[b]{20pc}\caption{\label{fig:fsrule}Dependence of the screening charge Z with the impurity potential $V_{0}$ for different interactions.}
                 \end{minipage}
                 \end{figure}

		The screening charge, defined by $Z=\sum_{i}|n_{i}-\bar{n}|$, depends on $V_{0}$ and g as is seen in Fig. \ref{fig:fsrule}. By changing $V_{0}$ this number increases upto $Z_{max}$ and then saturates 
		and even weakly decreases. We also see that by switching on the interaction, increasing g, the screening density is supressed. We found that the behaviour of Z is well described by
		the formula, 
		\begin{equation}
		  Z=Z_{max}(g) (1-e^ {-\frac{V_{0}}{t}}) ,
 		\end{equation}
              where $Z_{max}(g)$ decreases with g as is shown in the inset to Fig. \ref{fig:fsrule}.

\begin{figure}[h]
                 \includegraphics[width=20pc]{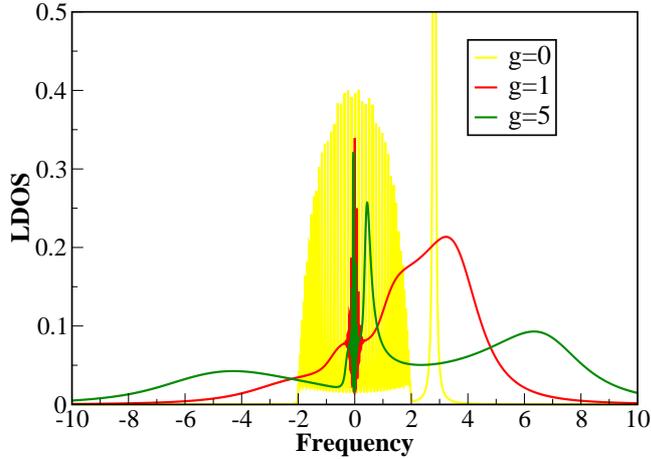}\hspace{2pc}%
                 \begin{minipage}[b]{14pc}\caption{\label{fig:specint}Local Density of States at the impurity site for different interactions. Redistribution of spectral weights occur
                 with increasing g. The antibound state $V_{0}=2$ peak for $g=0$ has finite width due to the numerical broadening parameter $\eta=0.008$.}
                 \end{minipage}
                 \end{figure}

                 In Fig. \ref{fig:specint} we present the Local Density of States (LDOS) at the impurity site for a 1D system. In case of $g=0$ we see the continuous band together with a peak
                   due to an antibound state at the energy of the order of $V_{0}$. The peak disappears with increasing g and in addition a pseudogap opens at the Fermi energy.

		 \subsection{2D interacting system}
\begin{figure}[h]
                  \begin{minipage}{14pc}
                  \includegraphics[width=10.5pc]{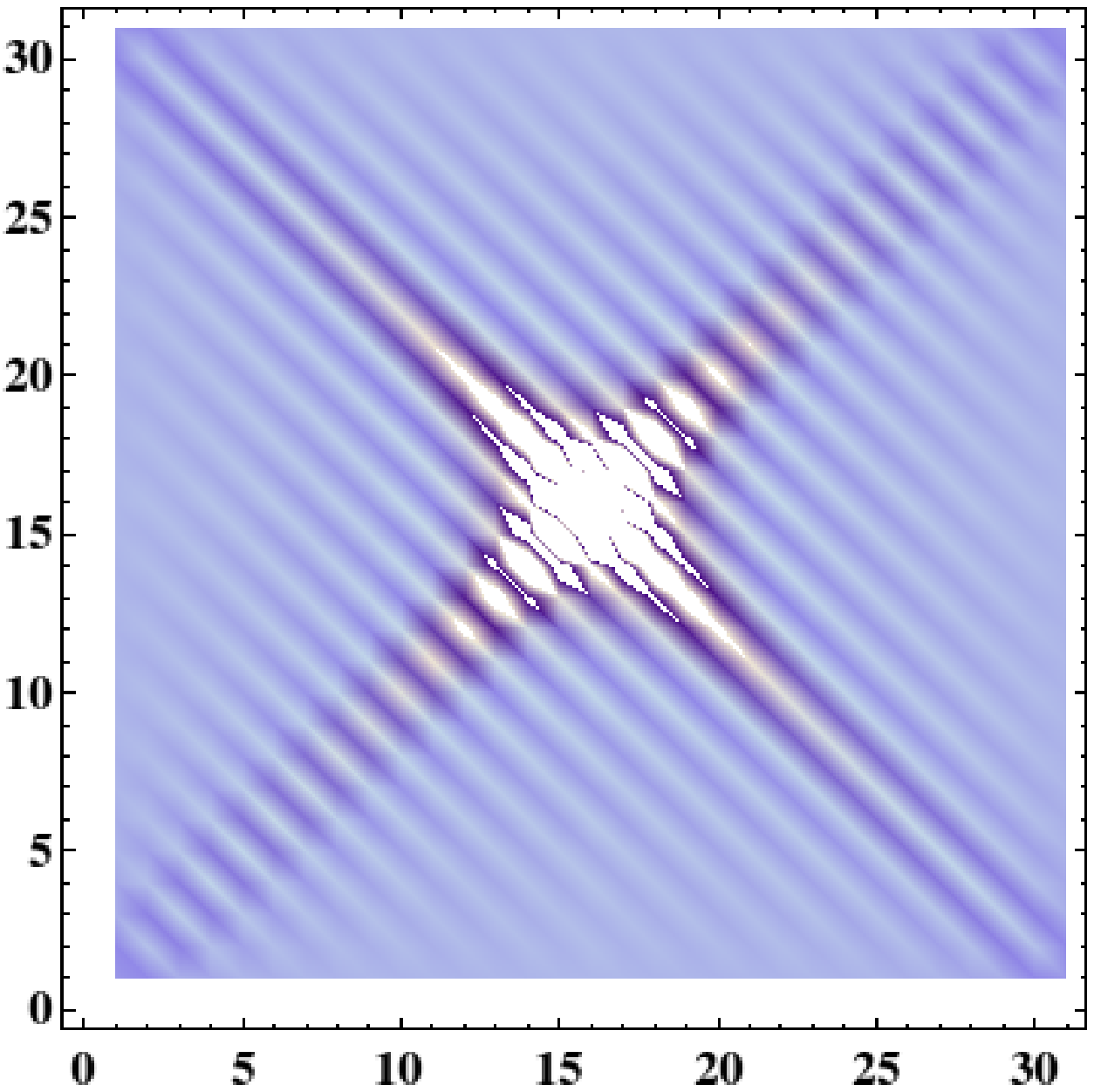}
                  \caption{\label{31contg0} Contour plot of the local occupation $n_{i}$ for the non-interacting (g=0) electrons on a 31 by 31 square lattice with impurity potential $V_{0}= -10$ 
                  located at the centre.}
                  \end{minipage}\hspace{2pc}%
                   \begin{minipage}{14pc}
                   \includegraphics[width=14pc]{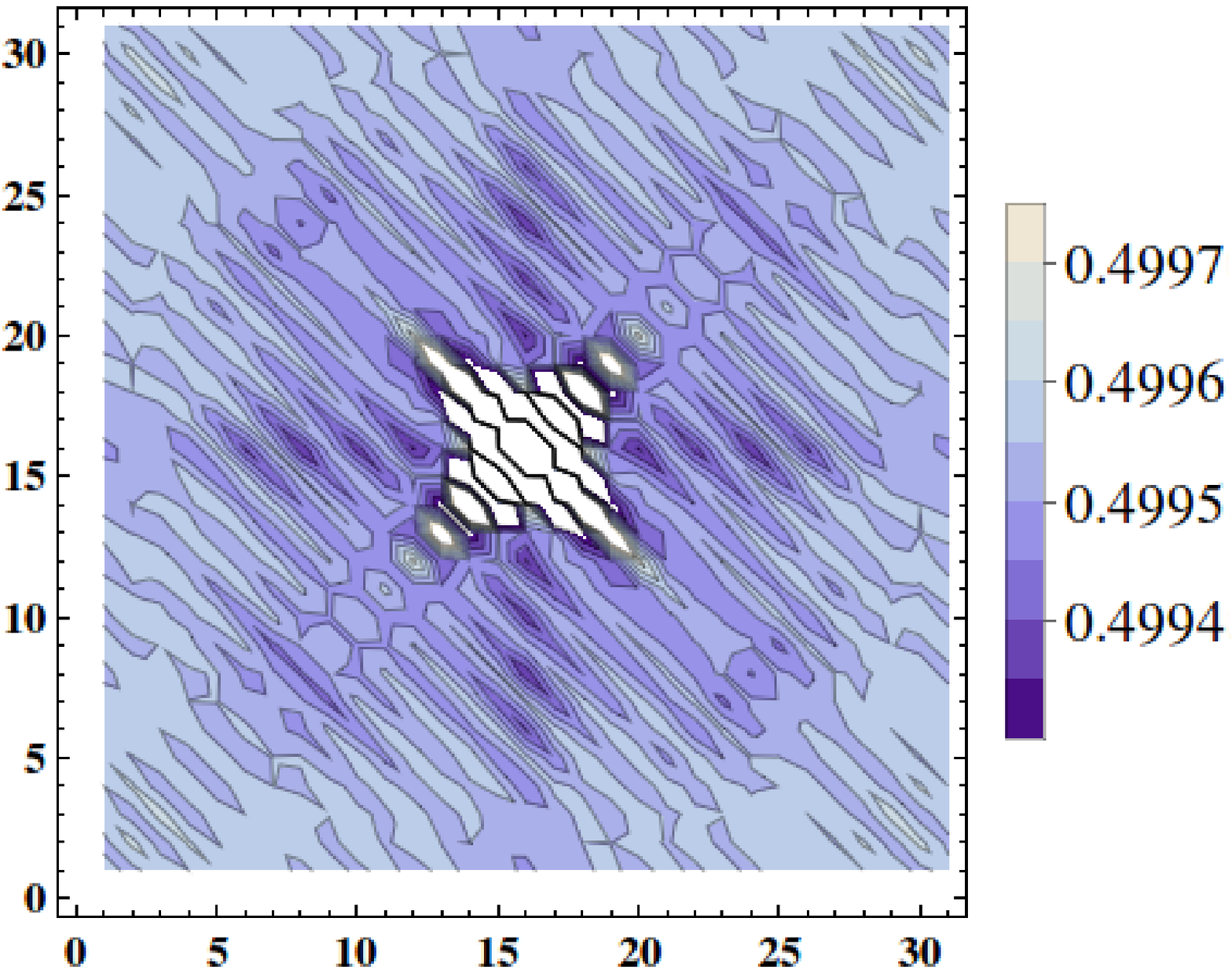}
                   \caption{\label{31contg3}Contour plot of the local occupation $n_{i}$ for the same system as in Fig. \ref{31contg0} but with g=3. Oscillations become more localised with increasing g.
                   }
                   \end{minipage} 
                    \end{figure}

                   The FO on a surface with non-interacting particles is shown in Fig. \ref{31contg0} where the system has 31 by 31 lattice sites and the impurity $V_{0}=-10$ is located at the centre. 
                   The symmetry of the square lattice is visible. In Fig. \ref{31contg3} we show the FO in the interacting case with $g=3$. The FO are more localised due to the interactions
                   and the oscillation pattern is richer.
		     
\section{Conclusion}
		    We have shown that in Fermi liquids the FO are influenced by electronic correlations.
		    In case of the one-dimensional systems the amplitudes of oscillations are suppressed with the interaction while the the period remains unchanged. In non-interacting 
		     and interacting systems the decay of oscillations follows the 1/r law. We observed that the saturation of the screening charge is supressed with increasing g.
		     We proposed a mathematical function describing the relation between the impurity potential and the screening charge. 
		     In case of 2D system with a single site impurity we observed that oscillations are more localised 
		     and  interference patterns appear with increasing g. This work provides us an initial knowledge for future studies
		     of FO in interacting electronic system using the full R-DMFT method. As mentioned earlier experimental studies on FO have mostly been conducted 
		     for non-interacting or weakly interacting system. 
		     It can be a challenging and interesting problem to test experimentally our numerical observations on correlated 
		     heavy fermionic system using STM \cite{aynajian2012visualizing, lee2009heavy}.

\subsection{Acknowledgments}
We thank  K. Makuch, J. Skolimowski, and D. Vollhardt for discussions. We acknowledge support from
the Foundation for Polish Science (FNP) through the TEAM/2010-6/2 project, co-financed by
the EU European Regional Development Fund. This research was also supported in part by the
Deutsche Forschungsgemeinschaft through TRR 80 (KB).

\end{document}